# Passive Lorentz Transformations with Spacetime Algebra

Carlos R. Paiva[a]

Instituto de Telecomunicações,

Department of Electrical and Computer Engineering, Instituto Superior Técnico

Av. Rovisco Pais, 1, 1049-001 Lisboa, Portugal

In special relativity, *spacetime algebra* (STA) provides a powerful and insightful approach to an invariant formulation of physics. However, in this geometric algebra of spacetime, relativistic physics is usually considered a misnomer: STA provides us with an invariant (coordinate-free) formulation and hence it seems that (passive) Lorentz transformations are misplaced. However, in order to interpret experimental results, it is important to relate invariant physical quantities to some reference frame. In this paper, passive Lorentz transformations are derived using STA but without Einstein's second postulate on the speed of light. This derivation sheds light on the physical background of STA and, at the same time, clarifies Lorentz transformations in the proper setting of (Minkowski) four-dimensional spacetime.

---

[a] e-mail: c.paiva@ieee.org

# I. Introduction

As argued by David Mermin, special relativity should be included in the high school curriculum.[1] In a first exposure to relativity one should stress the physical aspects without using the Lorentz transformations. At some point, nevertheless, it is inevitable to establish these transformations. However, there are two radically opposite ways of deriving them: either with or without Einstein's second postulate on the speed of light. One of the most famous derivations using the second approach is due to Lévy-Leblond.[2]

Apart from its intrinsic physical significance, Maxwell equations represent a landmark in physics that prompted some of the most important developments, both in physics and in mathematics, that took place ever since.[3] Special relativity is undoubtedly a consequence of those scientific advances – although, from an epistemological perspective, one should consider its foundations as independent from electromagnetism. On the other hand, it is possible to derive Maxwell equations using a metric-free approach by using the calculus of exterior differential forms.[4]

Special relativity suggested the following words from Hermann Minkowski: «Henceforth space by itself, and time by itself, are doomed to fade away into mere shadows, and only a kind of union of the two will preserve an independent reality.»[5] The *spacetime algebra* (STA) developed, among others, by David Hestenes is, from the author's viewpoint, the most adequate mathematical language to express that union.[6][7][8][9][10] However, when studying STA the Lorentz transformations are almost always assumed as a given. As far as the author knows, the derivation of the Lorentz transformation using STA can only be found in Chapter 9 of Ref. 7. Nevertheless, that derivation doesn't fit together with later approaches to STA as presented in Ref. 9 or in Chapter 5 of Ref. 10. In fact, STA provides an invariant (coordinate-free) formulation of physics through an elegant and concise manipulation of active Lorentz transformations, trying to explain *ab initio* that coordinates should be avoided. However, to overcome the coordinate "virus", it is important to be able to translate and dissolve passive Lorentz transformations in the fluidity and flexibility of STA, thereby bridging the gap between relativistic physics and proper spacetime physics. Furthermore, in order to interpret experimental results it is necessary to relate invariant physical quantities to some reference frame. In this paper the Lorentz transformation is derived without Einstein's second postulate on the speed of light using STA, i.e., working with vectors in a four-dimensional (flat) spacetime. The main purpose is to emphasize the physical background rather than a complete mathematical formulation, clarifying the four-dimensional setting of Lorentz transformations.

## II. The Geometric Product of Vectors

The elementary three-dimensional vector analysis developed by Gibbs is inappropriate to study special relativity. In fact, an event $a$ in spacetime belongs to a vector space $M$ with four dimensions and hence we cannot use the cross product of vectors, only defined in three dimensions, to multiply vectors in *spacetime*. Therefore, a new product between vectors should be defined when working with vectors in $M$. In an inertial (non-accelerating) frame $S$, the event (or vector) $a$ can de described by its four rectangular coordinates $(\kappa t, x, y, z) \in \mathbb{R}^4$ as follows:

$$a = (\kappa t)\mathrm{e}_0 + \mathbf{r}, \quad \mathbf{r} = x\mathrm{e}_1 + y\mathrm{e}_2 + z\mathrm{e}_3. \tag{1}$$

Here $\kappa$ is just an appropriate constant (with the dimension of velocity) that transforms units of time into units of length; no other meaning is ascribed *a priori* to it. The coordinates in $\mathbb{R}^4$ correspond to the relativistic viewpoint, whereas $a$ belongs to the so-called Minkowski spacetime (or manifold) $M$ that corresponds to proper spacetime physics. However, in another frame $\overline{S}$ moving away from $S$, the same event $a \in M$ is described by the new coordinates $(\kappa\overline{t}, \overline{x}, \overline{y}, \overline{z})$ in the new basis $(\mathrm{f}_0, \mathrm{f}_1, \mathrm{f}_2, \mathrm{f}_3)$:

$$a = (\kappa\overline{t})\mathrm{f}_0 + \overline{\mathbf{r}}, \quad \overline{\mathbf{r}} = \overline{x}\mathrm{f}_1 + \overline{y}\mathrm{f}_2 + \overline{z}\mathrm{f}_3. \tag{2}$$

A rest point $P$ in frame $S$ is described by the world line $a_P(t) = (\kappa t)\mathrm{e}_0 + \mathbf{r}_P$ where $\mathbf{r}_P$ is a constant vector ($t$ is the proper time of $S$). We always consider orthogonal bases, with $\mathrm{e}_\mu \cdot \mathrm{e}_\nu = 0$ as long as $\mu \neq \nu$. Therefore $\mathbf{r}_P$ is a spatial vector orthogonal to the time vector $\mathrm{e}_0$. Henceforth we always consider

$$\mathrm{e}_0^2 = \mathrm{e}_{\overline{0}}^2 = 1, \tag{3}$$

although the choice $\mathrm{e}_0^2 = \mathrm{e}_{\overline{0}}^2 = -1$ was also possible. Similarly, a rest point $Q$ in frame $\overline{S}$ is described by the world line $a_Q(\overline{t}) = (\kappa\overline{t})\mathrm{f}_0 + \overline{\mathbf{r}}_Q$ where $\overline{\mathbf{r}}_Q$ is also a constant vector ($\overline{t}$ is the proper time of $\overline{S}$). The world line of point $Q$, from the perspective of frame $S$, should be $a_Q(t) = (\kappa t)\mathrm{e}_0 + \mathbf{r}_Q(t)$. The proper velocity of point $P$ is $u_1 = da_P/dt = \kappa\mathrm{e}_0$, whereas the proper velocity of point $Q$ is $u_2 = da_Q/d\overline{t} = \kappa\mathrm{f}_0$. From Eqs. (3), one has $u_1^2 = u_2^2 = \kappa^2$. Moreover, as $u_2 = (da_Q/dt)(dt/d\overline{t})$,

$$u_2 = \gamma(u_1 + \mathbf{v}), \tag{4}$$

with $\gamma = dt/d\bar{t}$ and $\mathbf{v} = d\mathbf{r}_Q/dt$. Introducing also $\mathbf{v} = v\mathbf{e}_1 = \beta\kappa\mathbf{e}_1$ (with $\beta = v/\kappa$), where $\mathbf{e}_1$ is a unit vector with either $\mathbf{e}_1^2 = 1$ or $\mathbf{e}_1^2 = -1$ (to be determined later), then

$$\mathbf{f}_0 = \gamma(\mathbf{e}_0 + \beta\mathbf{e}_1). \tag{5}$$

When $\beta > 0$, the moving frame is receding (according to the observer's frame); when $\beta < 0$, the moving frame is approaching (also according to the observer's frame).

To proceed our derivation we introduce now the *geometric product* between spacetime vectors $(a,b) \mapsto ab$, with $a,b \in M$. We define this product as associative, (left and right) distributive over addition, and obeying to the so-called contraction $a^2 = aa \in \mathbb{R}$. It is this last property that distinguishes the geometric product from a general associative algebra. We do not force the contraction to be positive, just a real number. The geometric (or Clifford) product $ab$ can be decomposed into a symmetric inner product $a \cdot b$ and an antisymmetric outer product $a \wedge b$, as follows:

$$a \cdot b = \frac{1}{2}(ab + ba), \quad a \wedge b = \frac{1}{2}(ab - ba). \tag{6}$$

The inner product returns a scalar $a \cdot b = b \cdot a$, whereas the outer product returns a *bivector* $a \wedge b = -b \wedge a$. To show that $a \cdot b \in \mathbb{R}$ is easy: from $(a+b)^2 = a^2 + b^2 + ab + ba$ we have $2(a \cdot b) = (a+b)^2 - a^2 - b^2$, so that $a \cdot b$ is indeed a real number. Vectors can be pictured as directed line segments; bivectors, on the other hand, provide a means of encoding an oriented plane. Hence, the geometric product can be defined as a *multivector* that, according to Eq. (6), is given by

$$ab = a \cdot b + a \wedge b. \tag{7}$$

The sum (or multivector) in Eq. (7) looks strange at first: usually one should only add objects of the same type. In fact, the right-hand side of Eq. (7) should be viewed in the same way as the addition of a real and an imaginary number: the result is neither a real number nor a pure imaginary number – it is a graded sum forming a complex number. From Eq. (7), $ab = ba$ if and only if $a \wedge b = 0$ ($a$ and $b$ are parallel); $ab = -ba$ if and only if $a \cdot b = 0$ ($a$ and $b$ are orthogonal). Multivectors can be broken up into terms of different grade: the scalar part is assigned grade 0, the vector grade 1 and bivectors grade 2. We denote the projection onto terms of a chosen grade $r$ by $\langle \ \rangle_r$, so that in Eq. (7) we have $a \cdot b = \langle ab \rangle_0$ and $a \wedge b = \langle ab \rangle_2$. As

$$\begin{aligned}(a \wedge b)^2 &= (ab - a \cdot b)(a \cdot b - ba) \\ &= (a \cdot b)(ab + ba) - a^2 b^2 - (a \cdot b)^2\end{aligned} \tag{8}$$

we get

$$(a \wedge b)^2 = (a \cdot b)^2 - a^2 b^2 \tag{9}$$

and hence $(a \wedge b)^2 \in \mathbb{R}$.

Using the geometric product it is now possible to present Eqs. (4) and (5) as follows

$$u_2 = L^2 u_1, \tag{10}$$

$$L^2 = \gamma \left(1 + \beta \hat{B}\right), \tag{11}$$

$$\hat{B} = e_1 e_0, \tag{12}$$

with $u_2 = \kappa f_0$ and $u_1 = \kappa e_0$. One should note that $\hat{B} = e_1 e_0 = e_1 \wedge e_0$ as $e_1 \cdot e_0 = 0$.

Bivector $B = \beta \hat{B}$ completely characterizes the relative velocity between frames $S$ and $\bar{S}$ in spacetime. Hence, $u_2 u_1 = L^2 u_1^2 = L^2 \kappa^2$. Then

$$L^2 = f_0 e_0, \tag{13}$$

so that, according to Eq. (11), $f_0 \cdot e_0 = \gamma$ and $f_0 \wedge e_0 = \gamma \beta \hat{B}$. We call the transformation $L: S \to \bar{S}$, where $L^2$ is given by Eq. (13), a *boost* (one should note that $L^2$ is a multivector, not a simple bivector). Then, using Eq. (9), we get $(f_0 \wedge e_0)^2 = \gamma^2 \beta^2 \hat{B}^2 = \gamma^2 - 1$ from Eqs. (3). Hence,

$$\gamma = \frac{1}{\sqrt{1 - \beta^2 \hat{B}^2}}. \tag{14}$$

However, as $e_1 e_0 = -e_0 e_1$,

$$\hat{B}^2 = (e_1 e_0)(e_1 e_0) = -e_1^2 e_0^2 = -e_1^2 \tag{15}$$

and hence either $\hat{B}^2 = -1$ or $\hat{B}^2 = 1$. The reverse of a product of vectors is defined by

$$(ab)^\dagger = ba. \tag{16}$$

Therefore, from Eqs. (12) and (11), one has $\hat{B}^\dagger = e_0 e_1 = -\hat{B}$ and $(L^2)^\dagger = e_0 f_0 = \gamma(1 - \beta \hat{B})$, respectively. Accordingly, from Eq. (13), we obtain $e_0 = (L^2)^\dagger f_0$, or

$$e_0 = \gamma(f_0 - \beta f_1), \tag{17}$$

where, according to the principle of relativity,

$$f_1 = \hat{B} f_0 \quad \therefore \quad \hat{B} = e_1 e_0 = f_1 f_0. \tag{18}$$

Then, from Eqs. (5) and (17), we get $f_1 = \gamma\left[e_1 + e_0(\gamma^2 - 1)/\gamma^2\beta\right]$ or, using Eq. (14),

$$f_1 = \gamma\left(e_1 + \beta\hat{B}^2 e_0\right). \tag{19}$$

Using Eqs. (14) and (15), one gets indeed $\hat{B} = e_1 e_0 = f_1 f_0$, according to Eqs. (5) and (19). Hence the velocity of $\bar{S}$, from the perspective of an observer in $S$, is $\mathbf{v} = \beta\kappa e_1$. Likewise the velocity of $S$, from the perspective of an observer in $\bar{S}$, is $\mathbf{w} = -\beta\kappa f_1$ as $u_1 = \gamma(u_2 + \mathbf{w})$ in a similar way as Eq. (4) was derived. One should note that, in the Galilean limit, $\beta = 0$ and $\gamma = 1$ in Eq. (14) and hence we get, from Eq. (19), $f_1 = e_1$ and $\mathbf{w} = -\mathbf{v}$ as $\beta\kappa = v$ (although $\kappa = \infty$ and $\beta = 0$ in this limit). In summary: using Eqs. (5) and (19), we know how it is possible to obtain the pair $(f_0, f_1)$ in $\bar{S}$ from the pair $(e_0, e_1)$ in $S$, according to a Lorentz transformation $L: S \to \bar{S}$ corresponding to a relative motion characterized by bivector $B = \beta\hat{B}$, as shown in Fig. 1.

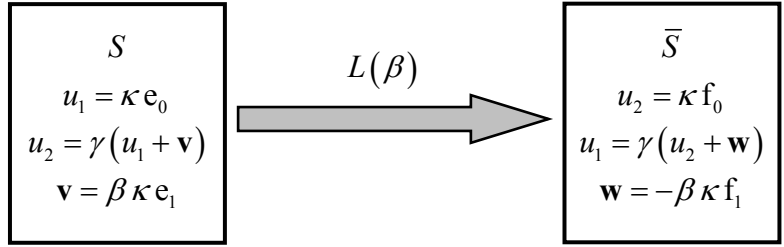

**Fig. 1. The Lorentz transformation (or boost) $L$ transforms the pair $(e_0, e_1)$ in frame $S$ into the pair $(f_0, f_1)$ in frame $\bar{S}$. In spacetime the relative motion between the two frames is completely characterized by bivector $B = \beta\hat{B}$, where $\hat{B} = e_1 e_0 = f_1 f_0$ and $\beta = v/\kappa$.**

## II. Passive Lorentz Transformations

To complete the derivation of the Lorentz transformation within the STA it is necessary to know, in Eqs. (15), whether $\hat{B}^2 = 1$ or $\hat{B}^2 = -1$.

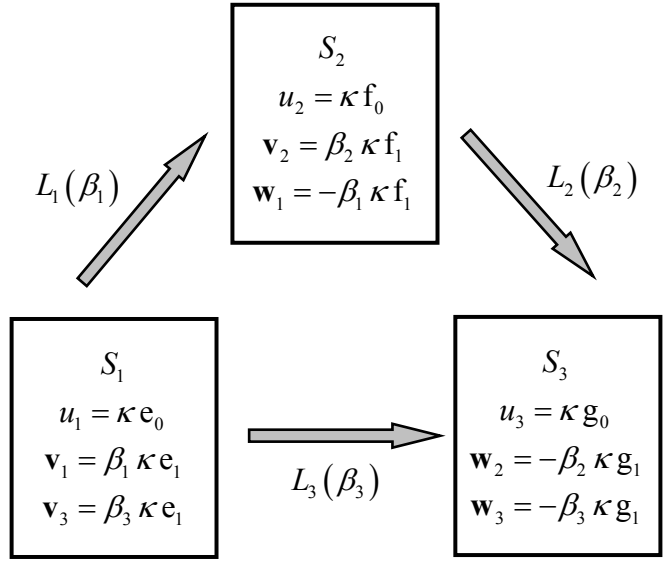

**Fig. 2.** The relative motion between frames $S_2$ and $S_1$ is characterized by bivector $B_1 = \beta_1 \hat{B}$, between frames $S_3$ and $S_2$ by $B_2 = \beta_2 \hat{B}$ and between $S_3$ and $S_1$ by $B_3 = \beta_3 \hat{B}$. One has $\hat{B} = e_1 e_0 = f_1 f_0 = g_1 g_0$.

Let us consider three frames: $S_1$, $S_2$ and $S_3$. A rest point in frame $S_k$ (with $k = 1, 2, 3$) has a proper velocity $u_k$ with: $u_1 = \kappa e_0$ in $S_1$; $u_2 = \kappa f_0$ in $S_2$; $u_3 = \kappa g_0$ in $S_3$. The relative motion between $S_2$ and $S_1$ is characterized by a relative velocity $B_1 = \beta_1 \hat{B}$; the relative motion between $S_3$ and $S_2$ by $B_2 = \beta_2 \hat{B}$; the relative motion between $S_3$ and $S_1$ by $B_3 = \beta_3 \hat{B}$. This situation is shown in Fig. 2 where, according to Eq. (11), $L_k^2 = \gamma_k \left(1 + \beta_k \hat{B}\right)$ with $\beta_k = v_k / \kappa$. For an observer in $S_1$, frames $S_2$ and $S_3$ are moving away with velocities $\mathbf{v}_1 = \beta_1 \kappa \mathbf{e}_1$ and $\mathbf{v}_3 = \beta_3 \kappa \mathbf{e}_1$, respectively; for an observer in $S_2$, frame $S_3$ is moving away with velocity $\mathbf{v}_2 = \beta_2 \kappa \mathbf{f}_1$. The physical problem is usually put as follows: given $\mathbf{v}_1$ and $\mathbf{v}_2$, one wishes to determine $\mathbf{v}_3$ in terms of $\mathbf{v}_1$ and $\mathbf{v}_2$. Only when all the relative motions take place along the same spatial direction, do we have, as in Fig. 2, $\mathbf{v}_1$ and $\mathbf{v}_2$ in $S_1$ both aligned along $\mathbf{e}_1$. Then, using Eqs. (5) and (19),

$$\mathbf{f}_0 = \gamma_1 \left(\mathbf{e}_0 + \beta_1 \mathbf{e}_1\right), \quad \mathbf{f}_1 = \gamma_1 \left(\mathbf{e}_1 + \beta_1 \hat{B}^2 \mathbf{e}_0\right), \tag{20}$$

$$\mathbf{g}_0 = \gamma_2 \left(\mathbf{f}_0 + \beta_2 \mathbf{f}_1\right) = \gamma_3 \left(\mathbf{e}_0 + \beta_3 \mathbf{e}_1\right). \tag{21}$$

Therefore, from Eqs. (20) and (21), we get

$$\beta_3 = \frac{\beta_1 + \beta_2}{1 + \beta_1 \beta_2 \hat{B}^2}. \tag{22}$$

For $\hat{B}^2 = -1$ and from Eq. (22), $\beta_3 = 2\beta / \left(1 - \beta^2\right)$ if $\beta_1 = \beta_2 = \beta$. Hence, $\beta_3 = 1$ for $\beta = \sqrt{2} - 1$ and $\beta_3 = \infty$ for $\beta = 1$; moreover, $\beta_3 < 0$ if $\beta > 1$. However, Eq. (22) is not acceptable on physical grounds: the composition of two velocities in the same spatial direction (e.g., $\beta_1 = 1$ and $\beta_2 = 2$) cannot lead to a velocity in the opposite direction $(\beta_3 = -3)$. Therefore, one should rule out the possibility of having $\hat{B}^2 = -1$ and always choose, according to Eq. (15),

$$\hat{B}^2 = -\mathbf{e}_1^2 = 1. \tag{23}$$

This means that Eq. (22) should reduce to

$$\beta_3 = \frac{\beta_1 + \beta_2}{1 + \beta_1 \beta_2} \quad \therefore \quad v_3 = \frac{v_1 + v_2}{1 + v_1 v_2 / \kappa^2}. \tag{24}$$

This last equation states the existence of a natural bound for velocities: there is an upper limit $\kappa$, to be determined through experience, for all particle velocities. Only when $\kappa = \infty$ do we recover the Galilean limit $v_3 = v_1 + v_2$. The fact that $\kappa < \infty$, however, can only be inferred from experimental evidence.

From Eqs. (23), the Lorentz transformation corresponding to Fig. 1, can be stated in the condensed form

$$e_0 \mapsto f_0 = L^2 e_0, \quad e_1 \mapsto f_1 = L^2 e_1, \tag{25}$$

according to Eqs. (5), (11) and (19). Furthermore, from Eqs. (14) and (23),

$$\gamma = \frac{1}{\sqrt{1 - \beta^2}}. \tag{26}$$

Then, as $|\beta| \leq 1$, it is possible to define a parameter $\zeta$, called *rapidity*, such that

$$\zeta = \tanh^{-1}(\beta) = \frac{1}{2} \ln\left(\frac{1+\beta}{1-\beta}\right). \tag{27}$$

Hence, using Eq. (11), we obtain

$$L^2 = \gamma\left(1 + \beta \hat{B}\right) = \cosh(\zeta) + \hat{B} \sinh(\zeta) \tag{28}$$

or, as $\hat{B}^2 = 1$ and

$$\exp(\zeta \hat{B}) = \sum_{n=0}^{\infty} \frac{1}{n!} (\zeta \hat{B})^n = \cosh(\zeta) + \hat{B} \sinh(\zeta), \tag{29}$$

$$L^2 = \exp(\zeta \hat{B}), \quad L = \exp\left(\frac{1}{2} \zeta \hat{B}\right), \tag{30}$$

$$L^{\dagger} = \exp\left(-\frac{1}{2}\zeta \hat{B}\right) \quad \therefore \quad LL^{\dagger} = 1. \tag{31}$$

Thus Eqs. (25) can be rewritten in the more symmetrical form

$$e_0 \mapsto f_0 = L e_0 L^{\dagger}, \quad e_1 \mapsto f_1 = L e_1 L^{\dagger}, \tag{32}$$

since $L e_0 = e_0 L^{\dagger}$ and $L f_1 = f_1 L^{\dagger}$ as $\hat{B} e_0 = -e_0 \hat{B} = e_1$ and $\hat{B} f_1 = -f_1 \hat{B} = f_0$, respectively.

All the necessary tools inside STA to derive the Lorentz transformations are now at one's disposal. Returning to Eqs. (1) and (2), it is now possible to rewrite them in the form

$$a = (\kappa t) e_0 + x e_1 + y e_2 + z e_3 = (\kappa \overline{t}) f_0 + \overline{x} f_1 + \overline{y} f_2 + \overline{z} f_3. \tag{33}$$

However, as vectors $e_2$ and $e_3$ are orthogonal to bivector $\hat{B} = e_1 e_0 = e_1 \wedge e_0$, they remain unchanged in a Lorentz transformation, i.e., $f_2 = e_2$ and $f_3 = e_3$. Hence, from Eqs. (33), we obtain

$$(\kappa t) e_0 + x e_1 = (\kappa \overline{t})(L^2 e_0) + \overline{x}(L^2 e_1). \tag{34}$$

Finally, using Eqs. (11), we get the passive Lorentz transformation

$$\begin{cases} \kappa t = \gamma(\kappa \overline{t} + \beta \overline{x}) \\ x = \gamma(\overline{x} + \beta \kappa \overline{t}) \\ y = \overline{y} \\ z = \overline{z} \end{cases}. \tag{35}$$

The inverse transformation $L^{\dagger}: \overline{S} \to S; (f_0, f_1) \mapsto (e_0, e_1)$ is obtained using Eqs. (32):

$$f_0 \mapsto e_0 = L^{\dagger} f_0 L, \quad f_1 \mapsto f_0 = L^{\dagger} f_1 L. \tag{36}$$

## IV. Concluding Remarks

In this paper the Lorentz transformation was derived without using Einstein's second postulate on the speed of light, thereby showing that special relativity is independent from electromagnetism. This approach is more epistemologically sound than the conventional one which relies on the fact that the speed of light $c$ (in a vacuum) is the same for all inertial observers. That, in Eqs. (24) and (35), one may write $\kappa = c$ is a result from electromagnetic theory and hence strictly outside the scope of special relativity. Ultimately, even if the photon mass was different from zero the special theory of relativity remained a consistent framework.

The derivation of the Lorentz transformation herein presented should not be considered within the context of a first exposure to relativity: the derivation refrained from glancing furtively at the doubters – something that cannot (and should not) be done at an introductory level. Nevertheless, the emphasis was not on the mathematical framework of STA but rather on the physical setting of Lorentz transformations in the four-dimensional Minkowski spacetime.

---

[1] N. David Mermin, *It's About Time: Understanding Einstein's Relativity* (Princeton University Press, Princeton, 2005).

[2] J-M. Lévy-Leblond, AJPIAS **44 (3)**, 271-277 (1976).